\documentclass[usegraphicx,usenatbib,useAMS]{mn2e}
\newcommand\lsim{\mathrel{\rlap{\lower4pt\hbox{\hskip1pt$\sim$}}
        \raise1pt\hbox{$<$}}}
\newcommand\gsim{\mathrel{\rlap{\lower4pt\hbox{\hskip1pt$\sim$}}
        \raise1pt\hbox{$>$}}}

\usepackage{amsmath}
\defcitealias{MH2007}{MH07}

\title[Suppression of HD--cooling in protogalactic gas clouds]{Suppression of HD--cooling in protogalactic gas clouds by Lyman-Werner radiation} 

\author[J. Wolcott-Green and Z. Haiman]{J. Wolcott-Green$^{1}$\thanks{E-mail:
    jemma@astro.columbia.edu; zoltan@astro.columbia.edu}
  and Z. Haiman $^{2}$\footnotemark[1]\\ $^{1}$Barnard College, 
  Columbia University, 3009 Broadway, New York, NY 10027\\ $^{2}$Department of
  Astronomy, Columbia University, 550 West 120th Street, New York, NY
  10027}

\begin{document}

\date{}

\pubyear{2010}

\maketitle

\begin{abstract}
  It has been shown that HD molecules can form efficiently in
  metal--free gas collapsing into massive protogalactic halos at high
  redshift. The resulting radiative cooling by HD can lower the gas
  temperature to that of the cosmic microwave background, $T_{\rm
  CMB}=2.7(1+z)$K, significantly below the temperature of a few
  $\times 100$K achievable via ${\rm H_2}$--cooling alone, and thus
  reduce the masses of the first generation of stars.
  Here we consider the suppression of HD--cooling by UV irradiation in
  the Lyman--Werner (LW) bands.  We include photo--dissociation of
  both ${\rm H_2}$ and HD, and explicitly compute the self--shielding
  and shielding of both molecules by neutral hydrogen, HI, as well 
  as the shielding of HD by ${\rm H_2}$.  
  We use a simplified dynamical collapse model, and follow the
  chemical and thermal evolution of the gas, in the presence of a UV
  background.
  We find that a LW flux of $J_{\rm crit,HD} \approx 10^{-22}{\rm erg
  \; cm^{-2}\; sr^{-1}\; s^{-1} \; Hz^{-1}}$ is able to suppress HD
  cooling and thus prevent collapsing primordial gas from reaching
  temperatures below $\sim 100$K. The main reason for the lack of HD
  cooling for $J>J_{\rm crit,HD}$ is the partial photo-dissociation of
  ${\rm H_2}$, which prevents the gas from reaching sufficiently low
  temperatures ($T<150$K) for HD to become the dominant coolant;
  direct HD photo--dissociation is unimportant except for a
  narrow range of fluxes and column densities.  Since the prevention
  of HD--cooling requires only partial ${\rm H_2}$
  photo--dissociation, the critical flux $J_{\rm crit,HD}$ is modest,
  and is below the UV background required to reionize the universe at
  $z \sim 10-20$.  We conclude that HD--cooling can reduce the
  masses of typical stars only in rare halos forming well before the
  epoch of reionization.

\end{abstract}


\section{Introduction}

The first generation of stars are believed to be much more massive
($\sim 100 {\rm M_\odot}$) than typical stars in stellar populations
in the low--redshift universe ($\sim 1 {\rm M_\odot}$;
\citealt{BCL02,ABN02}).  This has many important consequences in the
early universe, for reionization, metal--enrichment, the formation of
seed black holes at very early times, and the observability of
first-generation galaxies.

The high masses result from the thermodynamical properties of ${\rm
H_2}$, the main coolant in low--temperature gas with a primordial
composition.  In particular, ${\rm H_2}$--cooling becomes ineffective
at temperatures below $\sim 200$K.  HD molecules, can, in principle,
cool the gas to much lower temperatures, but until recently, the
abundance of HD in the early universe was believed to be too low for
it to be important.

It has recently been pointed out that significant HD can form in
metal--free gas, due to non-equilibrium chemistry, provided that the
gas has a large initial electron fraction.  This can occur, for
example, in ``fossil'' gas that was ionized by a short-lived massive
star, prior to it being extinguished, or in collisionally-ionized
halos with virial temperatures above $\approx 10^4$K.  It has been
shown that the resulting radiative cooling by HD can then lower the
gas temperature to values near that of the cosmic microwave
background, $T_{\rm CMB}=2.7(1+z)$K, i.e. to $\sim 30$K at $z\sim 10$.
This would decrease the expected masses of the stars that form in
ionized halos by a factor of $\sim 10$ below that which is possible if
HD-cooling is neglected \citep[e.g.][]{JB06}.
\footnote{Using three--dimensional simulations, \citet{MB08} found
  that HD lowers the expected Pop. III masses less dramatically, but
  still has an important effect.}  Thus, a second mode of star
  formation has been proposed, giving rise to Pop. III.2 stars
\footnote{Adopting the terminology suggested by \citet{BYHM09}.} that
can form as soon as a small number of Pop. III.1 stars have initiated
the epoch of reionization, and whose masses are only a few tens of
solar masses (\citealt{NU02}; \citealt{Mach05}; \citealt{NO05};
\citealt{JB06}, \citealt{Rip07}; \citealt{YOKH07}a; \citealt{YOH07}b).

These conclusions could potentially be revised, however, due to the
suppression of HD--cooling by UV irradiation of the gas
cloud. Although this possibility has been raised in the literature
(e.g. \citealt{JB06}; \citealt{YOH07}b), previous work has not
included a detailed treatment of of the impact of UV irradiation on
HD--cooling, including photo-dissociation of HD by radiation in its
Lyman and Werner (hereafter LW) bands, taking into account the
shielding that occurs in the optically thick regimes.
Such UV radiation will exist in the early universe, and can suppress
${\rm H_2}$--cooling in low--mass halos at high redshifts (e.g.,
\citealt{HRL97}).  {\em The main goal of this paper is to assess
whether HD--cooling can be similarly suppressed by UV radiation, and
to compute the critical UV flux for the HD-destruction}.

In order to do this, we perform ``one zone'' calculations with a
simplified density evolution, while following the gas--phase chemistry
and thermal evolution of the gas, including the impact of
${\rm H_2}$-- and HD--dissociating LW radiation.  In general, collapsing gas
clouds become optically thick to this radiation, so that the effects
of self--shielding are non-negligible.  Our treatment includes
self--shielding of HD and ${\rm H_2}$, shielding of both 
species by neutral hydrogen (HI), and shielding of HD by
${\rm H_2}$. We provide useful fitting formulae for these
shielding factors, analogous to the case of ${\rm H_2}$
self--shielding studied by \citet{DB96} (hereafter DB96).

The rest of this paper is organized as follows. In \S~\ref{sec:model}
we describe our chemical, thermal, and dynamical modeling.
\S~\ref{sec:results} presents our results on the critical flux required
to suppress HD--cooling, followed by a brief discussion of the
potential cosmological implications and primary uncertainties in
\S~\ref{sec:discussion}.  We summarize our main results and offer our
conclusions in \S~\ref{sec:conclusions}.  Throughout this paper, we
adopt the standard ${\rm \Lambda CDM}$ cosmological background model,
with the following parameters: ${\rm \Omega_{DM}=0.233,\;
\Omega_{b}=0.0462,\; \Omega_{\Lambda}=0.721,\; and}\; {\it h}= 0.701$ \citep{KDN09}.

\section{Model Description}
\label{sec:model}

The formation of HD occurs primarily through the following reaction
sequence (e.g. \citealt{GP02}):
\begin{equation}
{\rm H + e^- \rightarrow H^-} + {\it h\nu}
\label{eq:reaction1}
\end{equation}
\begin{equation}
{\rm H + H^- \rightarrow H_2 + e^-}
\label{eq:reaction2}
\end{equation}
\begin{equation}
{\rm D^+ + H_2 \rightarrow HD + H^+.}
\label{eq:reaction3}
\end{equation}
Thus, in order to form a significant abundance of HD, a large initial
electron fraction is required to catalyze the formation of ${\rm H_2}$
\citep[see, e.g.][and references therein]{JB06}.  In primordial gas
this can be achieved by photoionization (e.g. by short-lived Pop III.1
stars), or by collisional ionization (in sufficiently massive halos).
We model the first case by a constant, low-density gas, initially at a
density comparable that of the intergalactic medium (IGM) at high
redshift, $n\approx10^{-7}(1+z)^3 \; {\rm cm^{-3}}$, and temperature
$T\approx 10^4$ K.  We model the second case by a pre-imposed density
evolution obtained from the spherical collapse model.

\subsection{One-Zone Spherical Collapse Model}
\label{subsec:onezone}
 
We adopt the model for homologous spherical collapse that has been
used in several previous studies (e.g. \citealt{OSH08}; hereafter
OSH08).  This simple one-zone treatment prescribes the density
evolution of the baryonic and dark matter (DM) components of a
collapsing halo.  Both are initialized with zero velocity at the
turnaround redshift, set throughout this paper to $z=17$.  The density
of the in-falling gas evolves on the free-fall timescale and that of
the DM is given by a top-hat overdensity until virialization, after
which it remains constant at its virial value. Compressional heating
is included in the thermal model, along with the processes listed in
\S~\ref{sec:model}.2.

Unless stated otherwise, we take the radius of the cloud to be
$R_{\rm c} = \lambda_J/2$, where the Jeans length is given by
\begin{equation}
\lambda_{\rm J} = \sqrt{\frac{\pi k_B T\rm_{gas}}{G\rho\rm_{gas}\mu m_{\rm p}}},
\end{equation}
Here $k_B$ is Boltzmann's constant, $T\rm_{gas}$ is the gas
temperature, $\mu$ is the mean molecular weight, and $m\rm_p$ is the
mass of the proton.  Note that the size of the cloud is required, in
practice, only in our calculations of the self--shielding factors (see
below), in order to specify the column densities of ${\rm H_2}$, HD, and HI.  

While this model is a vast simplification of the physics of a
collapsing halo, it nonetheless has been shown to mimic the thermal
and chemical evolution seen in full three-dimensional hydrodynamical
simulations very well (see, e.g., \citealt{SBH10} -- hereafter SBH10
-- for a direct comparison).  The exception is the shock-heating that
occurs in the early stages of collapse and is not present in the
one-zone model, which prescribes a smooth ``free--fall''
evolution. For a detailed description of the spherical collapse model,
the reader is referred to the recent work by OSH08 and references
therein.

\subsection{Chemical and Thermal Model}
\label{subsec:chemistry}

We model a gas of primordial composition using a reaction network 
which comprises 47 gas-phase reactions amongst the 
following 14 chemical species (and photons): ${\rm H,~ H^+,~H^-,
~ He,~ He^+,~ He^{2+},~ H_2,~ H_2^+,~D,~D^+,~ D^-,~HD,}$\linebreak 
${\rm HD^+}$, and electrons.
Our choices for the selection of species and their initial abundances
are conventional \cite[see, e.g.,][]{GP98}, but we do not include any
lithium species or other potential coolants (e.g. ${\rm H_3^+}$), as
they contribute very little to the total cooling \citep[e.g.][]{GS09}
and are not important in the context of this paper.  

\subsubsection{Hydrogen and Helium Chemistry}
\label{subsubsec:Hchemistry}

The collisional rate coefficients for reactions among hydrogen
and helium species only, and cross-sections for photo-ionization, 
are taken from the recent compilation by SBH10. However, 
the rate for ${\rm H_2}$ photo-dissociation ($k_{28}$ in the
aforementioned compilation) is modified to $k_{\rm diss,H_2}=
1.39 \times 10^{-12} \times \beta \times f_{\rm shield}$
in order to match the optically thin rate we calculate
(see \S~\ref{subsec:opthin}). The total shielding 
is parameterized by a shield factor, $f_{\rm shield}$ 
(see below), and the rate is normalized by the 
parameter $\beta$, as described by in Appendix A 
of OM01, which specifies the intensity of blackbody 
radiation at the average LW band energy ($12.4$ eV) relative 
to that at the Lyman limit ($13.6$ eV). For the two spectral
types we consider (described in \S~\ref{sec:results}), 
$\beta = 3$ for the T4-type and $\beta = 0.9$ for the T5-type.

\subsubsection{Deuterium Chemistry}
\label{subsubsec:Dchemistry}

The chemical network includes 19 reactions involving the five 
deuterium species, for which we use the collisional rate 
coefficients from the compilation by \citet{NU02}. However, 
we replace the rates D2, D3, D7, and D9 given therein 
(as referenced in the source) with the corresponding
updated rates from \citet{Savin02} (for the charge exchange
reactions, D2 and D3) and \citet{GP02} (for D7 and D9). 
The HD photo--dissociation rate is given in 
\S~\ref{subsec:opthin} and is normalized with the 
$\beta$ parameter in the same manner as described 
above for the ${\rm H_2}$ photo--dissociation rate. 

We take the cosmological D/H ratio to be
4$\times 10^{-5}$ by number, following recent studies on HD--cooling
(e.g.~\citealt{JB06},~\citealt{YOKH07}) and inspired by the 
model of \citet{GP98}, which provides a value of D/H $=4.3\times 10^{-5}$.
This adopted value is likely overgenerous for
the primordial deuterium abundance, however, in light 
of recent observations, which place estimates of D/H at
$2.78 ^{+0.44}_{-0.38}\times 10^{-5}$ \citep{Tytler03} 
and $2.82 ^{+0.27}_{-0.25}\times 10^{-5}$ \citep{Omeara06}. 
However, decreasing the initial deuterium abundance in our models 
leads to less robust HD--cooling, and so only serves to strengthen 
our central conclusion that metal-free gas is unlikely to be cooled,
by HD, to temperatures close to $T_{\rm CMB}$.

In \S~\ref{sec:discussion}, we discuss recently updated rate 
coefficients for some of the most important reactions, 
and how their implementation affects our results.

\subsubsection{Thermal Model}
\label{subsubsec:heatcool}

The following processes are included in the net cooling rate: 
collisional excitation and 
ionization (of H, He, and ${\rm He^+}$), recombination (to H, He, and
${\rm He^+}$), dielectric recombination (to He), Bremsstrahlung,
Compton cooling,\footnote{Numerical expressions for these cooling
processes can be found in, e.g., \citet{HRL96}.} and molecular cooling
by ${\rm H_2}$ and HD.  In practice, the last two processes, as well as
collisional excitation of HI, dominate in our calculations.
We adopt the expression provided by \citet{GP98} for ${\rm H_2}$
cooling.  In the fossil gas case, the HD--cooling rate is calculated
using the analytic fit for low densities ($n \lsim 10^3 {\rm
  cm^{-3}}$) given by equation (5) in \citet{Lipov05}. In the
spherical collapse runs, we adopt the lengthier polynomial fit
(equation 4 in the same source), which is accurate for gas densities
$n \lsim 10^8\; {\rm cm^{-3}}$.

We note that $T_{\rm {CMB}}$ is a ``temperature floor,'' below which
gas cannot cool radiatively; if the gas temperature were below $T_{\rm
  {CMB}}$, interaction with photons in the roto--vibrational bands
would heat, rather than cool the gas. In order to mimic this behavior,
we multiply $\Lambda_{\rm H_2}$ and $\Lambda_{\rm HD}$ by a correction
factor $\left(T - T_{\rm {CMB}}\right)/ \left(T + T_{\rm
    CMB}\right)$. This ensures that cooling is shut-off as the
temperature approaches $T_{\rm CMB}$ from above (whereas the
correction becomes negligible when $T\gg T_{\rm CMB}$; see
\citealt{HRL96} and \citealt{JB06} for a somewhat more accurate
approach).

Our thermal model includes heating from photo-detachment of ${\rm  H^-}$, 
high energy electrons resulting from photo-ionization of helium 
\citep[see, e.g., Appendix B in][]{HRL96}, as well as compressional heating in 
the model of adiabatic collapse (see OSH08 and references therein for more details).  
In practice, the latter dominates in the regime of our calculations.

In order to follow the coupled chemical and thermal evolution of the
gas we use the Livermore solver LSODAR to solve the stiff equations.

\subsection{HD and ${\rm H_2}$ Photo-dissociation in the Optically Thin Limit}
\label{subsec:opthin}

HD and ${\rm H_2}$ can be dissociated by photons with energies in the
range 11.2-13.6 eV, to which the universe is largely transparent even
before the IGM is reionized. Although both HD and ${\rm H_2}$ have LW
lines above 13.6 eV, we do not include photons above this energy,
because they will have been absorbed by neutral hydrogen elsewhere in
the IGM prior to reionization.  
Here we describe the details of the calculation for HD photo-dissociation; 
however, the calculation for ${\rm H_2}$ is entirely analogous, 
so the following  applies equally well to both molecules.

Excitation of the HD molecule to its $B^1 \sum_u^+$ and $C^1 \Pi_u$
electronic states and subsequent radiative decay leads to dissociation
when the system decays to the vibrational continuum of the ground
state, rather than back to a bound state.  Here we discuss
the optically thin case, in which the processing of the LW
spectrum by HD itself (as well as by ${\rm H_2}$ and HI) is assumed to
be negligible.  The dissociation rate for molecules initially
in the electronic ground state with vibrational and 
rotational quantum numbers (${\it  v,J}$) is given by:
\begin{equation}
k_{{\rm diss,}{\it v,J}} = \sum_{\it v',J'} \zeta_{\it v,J,v',J'} {\it
f}_{{\rm diss,}{\it v',J'}}
\end{equation}
where $\it{f}\rm_{diss,\it{v',J'}}$ is the dissociation probability
from the excited state (${\it v',J'}$) and the pumping rate is given by
\begin{equation}
\zeta_{\it{v,J,v',J'}} = \int_{\it{\nu_{th}}}^{\infty}4\pi\sigma_{
%\nu}\left(v,J,v',J'\right)\frac{J_{\rm \nu}}{h_P \nu}{\rm d}\nu.
\nu}\frac{J_{\rm \nu}}{h_P \nu}{\rm d}\nu.
\end{equation}
Here $\sigma_{\nu}$ is the frequency--dependent cross-section of
a given transition, $h_P$ is
Planck's constant, and the specific intensity just below 13.6 eV is
hereafter normalized as $J_{\nu}= J_{21} \times [10^{-21}\,{\rm erg \;
  cm^{-2}\; sr^{-1}\; s^{-1} \; Hz^{-1}]}$.  As mentioned above, in
our model there is a sharp cut off in the radiation spectrum above
13.6 eV; the lower limit, ${\it \nu_{th}}$, is the frequency threshold,
corresponding to the longest--wavelength photons included, 
$\lambda {\rm \sim 1105\; \AA}$.

In principle, the total dissociation rate also depends on the level
populations of the molecule, which in turn depend on the incident radiation
field as well as the temperature and density of the gas; thus, the
total dissociation rate should be:
\begin{equation}
k_{\rm {diss,tot}} = \sum_{\it {v,J}}k_{\rm {diss,{\it{v,J}}}} f_{v,J},
\label{eq:kdisstot}
\end{equation}
where $f{\it_{v,J}}$ is the fraction of molecules initially in the
ro-vibrational state denoted by ${\it {v,J}}$. For simplicity, we
assume that all HD and ${\rm H_2}$ molecules are initially 
in the ground state (i.e. ${\it f_{v=0,J=0}} = 1$). This is a 
reasonable approximation for low gas densities, at which the 
populations of higher ro-vibrational states are very small. 
However, the level populations of ${\rm H_2}$ and HD reach their
values at local thermodynamic equilibrium when gas densities rise to 
$n\gsim 10^4~{\rm cm^{-3}}$, and $n\gsim 10^6~{\rm cm^{-3}}$ 
respectively. In \S~\ref{sec:discussion}, we discuss the 
differences in the dissociation rates if both molecules are 
assumed to be in LTE, and how this impacts the results discussed below.

We include 28 discreet spectral lines of HD and 25 of ${\rm H_2}$,
all involving transitions from from the ground electronic state 
${\rm X^1 \sum_{\it g}^+}$ to the $B^1 \sum_u^+$ and $C^1 
\Pi_u^+$ excited states. We use the necessary data for the 
Lyman and Werner bands of HD provided by \citet{AR06}. 
For those of ${\rm H_2}$, the relevant data were taken from 
\citet{ARLa93}, and \citet{ARLb93}. We use the updated 
dissociation fractions for ${\rm H_2}$ in \citet{ARD00}.
The numerical wavelength resolution in the calculations ($\Delta
\lambda = 5.8 \times 10^{-5}{\rm \AA}$ at the lowest temperatures) 
is sufficient to resolve the Voigt profile of each line and explicitly
account for overlap of the Lorentz wings.  We find the 
following photo--dissociation rates in the optically thin limit:
$k {\rm_{{diss,HD}} =1.55 \times 10^9}J_{\bar{\nu}} \;{\rm s^{-1}}$, 
and $k {\rm_{{diss,H_2}} =1.39 \times 10^9}J_{\bar{\nu}} \;{\rm s^{-1}}$,
in excellent agreement with those found previously by \citet{GJ07}:
$k {\rm_{{diss,HD}} =1.5 \times 10^9}J_{{\bar\nu}} \;{\rm s^{-1}}$,
and $k {\rm_{{diss,H_2}} =1.38 \times 10^9}J_{\bar{\nu}} \;{\rm s^{-1}}$.
Here $J_{\bar{\nu}}$ denotes the intensity at the average LW band
of HD and ${\rm H_2}$, with energy 12.4 eV, as discussed above.

\subsection{Self-Shielding of HD and ${\rm H_2}$}
\label{subsec:selfshield}

\begin{figure}
    \includegraphics[width=3.3in]{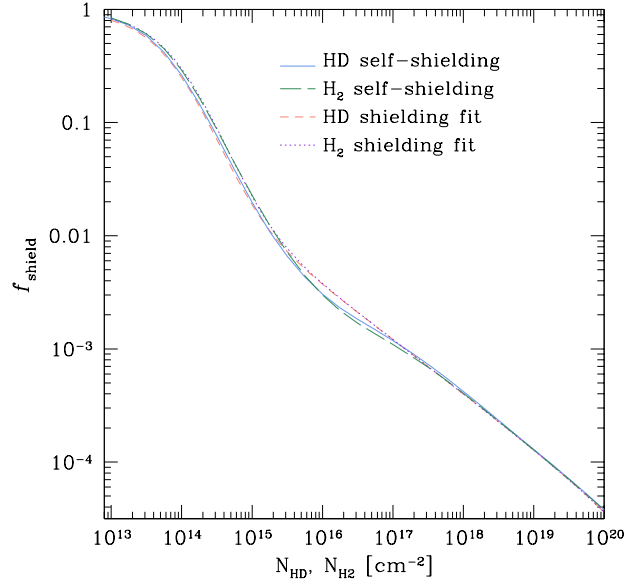}
    \caption{Solid (blue) and broken (green) curves show the 
      numerically calculated self--shielding factors for HD 
      and ${\rm H_2}$ respectively; these are defined as the ratio of 
      the dissociation rate at a given column density and 
      the optically thin dissociation rate, and are shown as
      functions of the respective column densities.
      Dashed (orange) and dotted (magenta) curves show the
      values obtained by using the fitting formulae from 
      equation~\ref{eq:selfshield} for HD and ${\rm H_2}$ 
      self--shielding, respectively (the slight difference here
      arises only because of the different masses of the two
      molecules, which modifies the Doppler parameter).  
      All are shown at $T=200$ K.}
\label{fig:selfshield}
\end{figure}

When sufficiently high column densities of HD or ${\rm H_2}$ build up,
($N_{\rm HD},N_{\rm H_2} \gsim 10^{13}\; {\rm cm^{-2}}$), the LW bands
become optically thick and the rates of photo--dissociation are
suppressed. We parameterize this effect by a shield factor, $f_{\rm
shield}$, akin to that given by DB96 in their study of 
${\rm H_2}$ self-shielding.  
In particular, $f_{\rm shield,HD} \equiv k_{\rm diss,HD}(N_{\rm
HD})/k_{\rm diss,HD}(N_{\rm HD}=0)$ where $k_{\rm diss,HD}(N_{\rm
HD}=0)$ is the dissociation rate in the optically thin limit
(equation~\ref{eq:kdisstot}), and the shield factor for ${\rm H_2}$,
$f_{\rm shield,H_2}$, is analogously defined.
Our treatment of ${\rm H_2}$ self--shielding differs from the
previous study by DB96 in that we assume all ${\rm H_2}$ is in the 
ro-vibrational ground state (as described above), while the latter
used a model allowing for populations in higher ro-vibrational 
levels due to collisional excitation and ``UV pumping'' 
by the incident radiation field. Nonetheless, we find that a 
good analytical fit for both ${\rm H_2}$ and HD self--shielding 
is provided by the same functional form as equation (37) for
$f_{\rm shield,H_2}$ in DB96. We also find that the self--shielding 
behavior of the two molecules is nearly identical, 
(see Figure~\ref{fig:selfshield}), which might be expected  
on the basis of the similarity in their electronic structures \footnote{
The similarities in the line strengths of ${\rm H_2}$ and HD, 
quantified by the product of the oscillator strength and dissociation
fraction, can be seen in Figure~\ref{fig:linelist} below.}; 
thus, we use the following fitting formula for both 
$f_{\rm shield,H_2} \left(N_{\rm H_2}, T\right)$ 
and $f_{\rm shield,HD}\left(N_{\rm HD}, T\right)$:
\begin{multline}
f_{\rm shield}\left(N, T\right) = 
\frac{0.9379}{\left(1 + {\rm x}/{\rm D_5}\right)^{1.879}} 
+ \frac{0.03465}{\left(1 + {\rm x}\right)^{0.473}}\\
\times \exp\left[-2.293 \times 10^{-4} \left( 1 + {\rm x}\right)^{0.5}\right],
\label{eq:selfshield}
\end{multline}
where ${\rm x} \equiv N/ 8.465 \times 10^{13} {\rm cm^{-2}}$,
$N$ is the column density of the self-shielding species,
 ${\rm D_5 \equiv b_{D}/ 10^5 {\rm cm~s^{-1}}}$, 
and the Doppler broadening parameter, ${\rm b_{D}}$, 
depends on the mass of the molecule (which accounts for the 
slight difference in the self-shielding formula for the
two molecules), as well as the temperature.

\subsection{Shielding by HI and Mutual Shielding of  ${\rm H_2}$ and HD}
\label{subsec:shielding}

In addition to self-shielding, HD and ${\rm H_2}$ can also 
shield each another, and both can be shielded by HI, 
which has absorption lines in the range 11.2-13.6 eV; 
thus, suppression of the photo-dissociation rates 
depend on the relative strengths and positions of the 
HD, ${\rm H_2}$, and HI lines, as well as the column densities
of each species, $N_{\rm HD}$, $N_{\rm H_2}$, and $N_{\rm HI}$.

\begin{figure}
\includegraphics[width=3.3in]{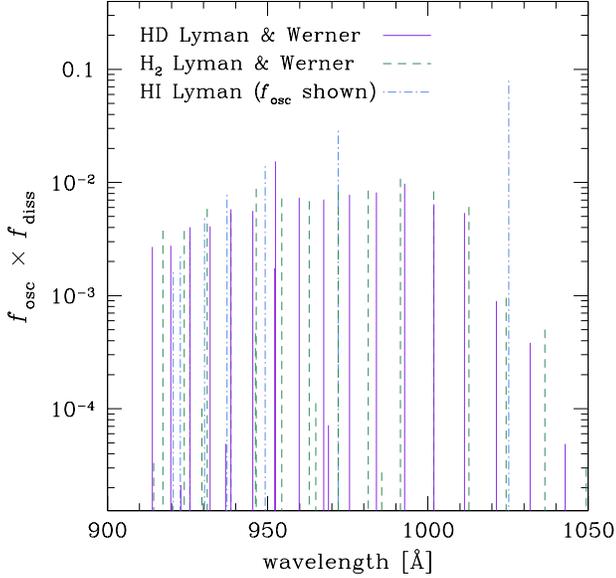}
    \caption{The figure illustrates the wavelengths and strengths of
      the relevant LW lines of ${\rm H_2}$ and HD, as well as the
      Lyman series of HI.  The solid (purple) and dashed (green) lines
      indicate the product of the oscillator strength and dissociation
      fraction for HD and ${\rm H_2}$ respectively, drawn at the position
      of each line center. The wavelengths and
      oscillator strengths of the hydrogen Lyman series are similarly shown
      by the blue dash-dot lines (a dissociation fraction in
      this case is not applicable). At high HI column densities, the 
      HD lines that dominate the dissociation rate are those at
      at $\sim 960,\sim 980$, and $\sim 990$ \AA, and the dominant 
      contributions to the ${\rm H_2}$--dissociation rate are made by
      the absorption lines at $\sim 945,\sim 980$, and $\sim 990$ \AA.}
\label{fig:linelist}
\end{figure}

We include the first nine Lyman HI lines in the relevant wavelength range; 
while the line center of the Ly$\alpha$ line is outside this range, we
nevertheless include it, as its contribution to the shielding becomes
important due to line broadening at high HI column densities.  For
illustration, in Figure~\ref{fig:linelist} we show the positions and
strengths of the most significant lines of each species.

\subsubsection{Shielding of HD by $H_2$ and HI}
\label{subsubsec:HDshield}

Taking shielding into account, the contribution to the 
HD--dissociation rate for a particular line at frequency 
$\nu$ becomes:
\begin{equation}
  k_{\rm diss,HD,\nu}(N_{\rm HD}) = k_{\rm diss,HD,\nu}(N_{\rm HD}=0)
  \;{\rm \exp(-\tau_{\nu})},
\end{equation}
where the optical depth is given by
\begin{equation}
  \tau_{\nu}=\sigma_{\rm HD,\nu} N_{\rm HD} + \sigma_{\rm HI,\nu}
  N_{\rm HI} + \sigma_{\rm H_2,\nu} N_{\rm H_2}.
\label{eq:tau_nu}
\end{equation}

\begin{figure}
    \includegraphics[width=3.3in]{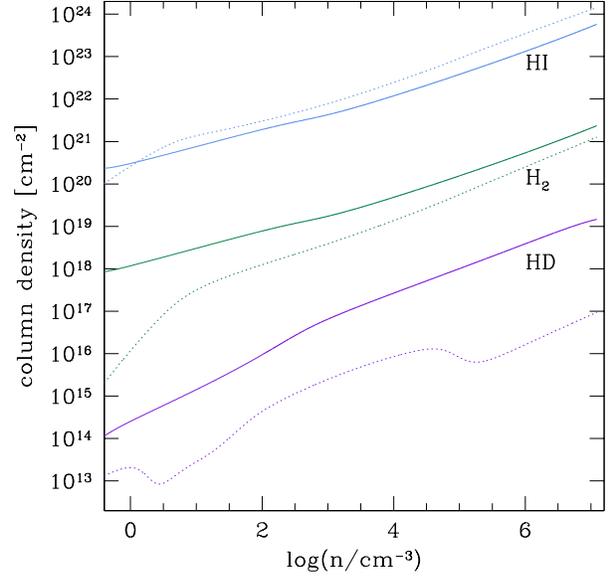}
    \caption{Column densities reached in the spherical collapse runs
      for halos with virial temperatures above and below $10^4$K,
      shown by solid and dotted lines respectively.  Both runs assume
      no background flux ($J=0$).}
\label{fig:columns}
\end{figure}

While HD self-shielding becomes important for $N_{\rm HD}\gsim
10^{13}\; {\rm cm^{-2}}$, the offsets in the wavelength of the
neighboring absorption lines (typically of order $\sim$\AA) prevent
${\rm H_2}$ and HI from effectively shielding HD until their column
densities are very high, $N_{\rm H_2}\gsim 10^{20}\; {\rm cm^{-2}}$,
and $N_{\rm HI}\gsim 10^{23}\; {\rm cm^{-2} }$. At these critical
densities, which are essentially independent of $N_{\rm HD}$, the
${\rm H_2}$ lines start to significantly overlap and the optical depth
due to ${\rm H_2}$-shielding is $\sim$ a few at all wavelengths.  In
Figure~\ref{fig:columns}, we show the evolution of all three column
densities, as a function of the particle number density, in our
one-zone collapse runs with $J=0$
for halos with virial temperatures both above and below $10^4$K (see
below).  This figure shows that all three column densities reach the
values where shielding becomes efficient, and thus a full calculation
of the combined self-shielding is warranted.

We provide a fitting formula to model the total shielding of HD, 
$f_{\rm shield, HD}= f_{\rm shield, HD}(N_{\rm HD},
N_{\rm H_2},N_{\rm HI},T)$ in 
Table~1 and equations~\ref{eq:shieldproduct},  and
\ref{eq:shield2} below, which is accurate to within a factor of two
over a wide range of column densities, i.e. up to $N_{\rm HD}\approx
10^{20}\; {\rm cm^{-2}},\; N_{\rm H_2}\approx 10^{22} \; {\rm
  cm^{-2}},\; N_{\rm HI}\approx 10^{24} \; {\rm cm^{-2}}$ and gas
temperatures up to $\approx 10^3$ K (HD--cooling is unimportant at
temperatures above this value in any case):

\begin{equation}
f_{\rm shield,HD} = f_{\rm shield}\left(N_{\rm HD},T\right) 
\times f_1\left(N_{\rm HI}\right) \times f_2\left(N_{\rm H_2}\right)
\label{eq:shieldproduct}
\end{equation}
\begin{equation} 
f_i = \frac{1}{\left(1 + {\rm x}_i\right)^{{\rm \alpha}_i}}
\times \exp\left({\rm-\beta}_i\;{\rm x}_i\right).
\label{eq:shield2}
\end{equation}
Here ${\rm x}_i \equiv N_i/\gamma_i$ and the index $i$ 
takes the value $i=1$ or 2 to denote the relevant quantity 
for HI or ${\rm H_2}$ respectively. 
The coefficients $\alpha$, $\beta$, and $\gamma$ are given in
Table~1.
\begin{table}
\begin{center}
  \caption{Coefficients for the fitting formula for $f_{\rm shield,
      HD}(N_{\rm HD},N_{\rm H_2},N_{\rm HI},T)$, representing the
    total combined shielding factor of HD, including shielding 
    by ${\rm H_2}$ and HI
    (equations~\ref{eq:shieldproduct} and \ref{eq:shield2}).
    The analytic fit for self-shielding is given in 
    equation~\ref{eq:selfshield}.}
  \label{tbl:fitform}
  \begin{tabular*}{0.475\textwidth}{@{\extracolsep{\fill}} l l l l }
    \hline\hline
    Species & $\alpha$ & $\beta$ & $\gamma$ (cm$^{-2}$)\\
    \hline\\
    1. {\rm HI} & 1.620 &  0.149 & $2.848 \times 10^{23}$\\
    2. ${\rm H_2}$ & 0.238 & 0.00520 & $2.339 \times 10^{19}$\\
    \hline
  \end{tabular*}
\end{center}
\label{table:fits}
\end{table}

This fit is described by a product of three separate functions, $f_{\rm
  shield, HD}(N_{\rm HD},T),\; f_{\rm shield, H_2}(N_{\rm H_2}),$ and
$f_{\rm shield, HI}(N_{\rm HI})$, which represent the shielding of HD
due to each of the three species alone (eq.~\ref{eq:shieldproduct} above).  
Note that since the ${\rm H_2}$ and HI lines shield HD by
their Lorentz wings, rather than their thermal cores, these factors
(unlike HD self-shielding) do not depend on temperature.

In general, one does not expect that the combined shielding factor is separable into a
simple product of the three individual shielding factors.  The full
expression for $f_{\rm shield, HD}(N_{\rm HD},N_{\rm H2},N_{\rm
  HI},T)$ is a sum over all of the individual cross-sections of the HD
lines, each suppressed by the frequency-dependent total optical depth
(eq.~\ref{eq:tau_nu}), divided by the optically thin rate.  
However, we have found that in practice, when suppression by nearby HI 
lines is negligible ($N_{\rm HI} \lsim 10^{23}~{\rm cm^{-2}}$),
one HD line is much stronger than all others (at ${\rm \approx 
950 \AA}$, see Figure~\ref{fig:linelist}). Since this single line
dominates the dissociation rate, the shield factor reduces to the 
simple product (eq.~\ref{eq:shieldproduct} above). 
In the regime of relatively strong HI shielding 
($N_{\rm HI} \approx 10^{24}~{\rm cm^{-2}}$), a few HD 
Lyman lines together dominate the dissociation rate. 
However, we find that the product $\sigma_\lambda\times 
f_{\rm diss,\lambda}$ for these lines (at $\sim$960, 
$\sim$980, and $\sim$990 \AA) are similar.  
If we approximate that these lines have identical 
strengths, the total shielding factor again reduces to the 
simple product in equation~\ref{eq:shieldproduct}
above. Because these line strengths are not precisely equal,
the largest discrepancies in the product formula and `true'
shielding behavior are seen at $N_{\rm HI} = 10^{24}~{\rm cm^{-2}}$.
However, in general, we find that this simple product is
accurate, to within a factor of $\sim$two, at the low temperatures
($T\lsim 200$ K) and the high column densities of interest.

\begin{figure}
    \includegraphics[height=2.8in,width=3.4in]{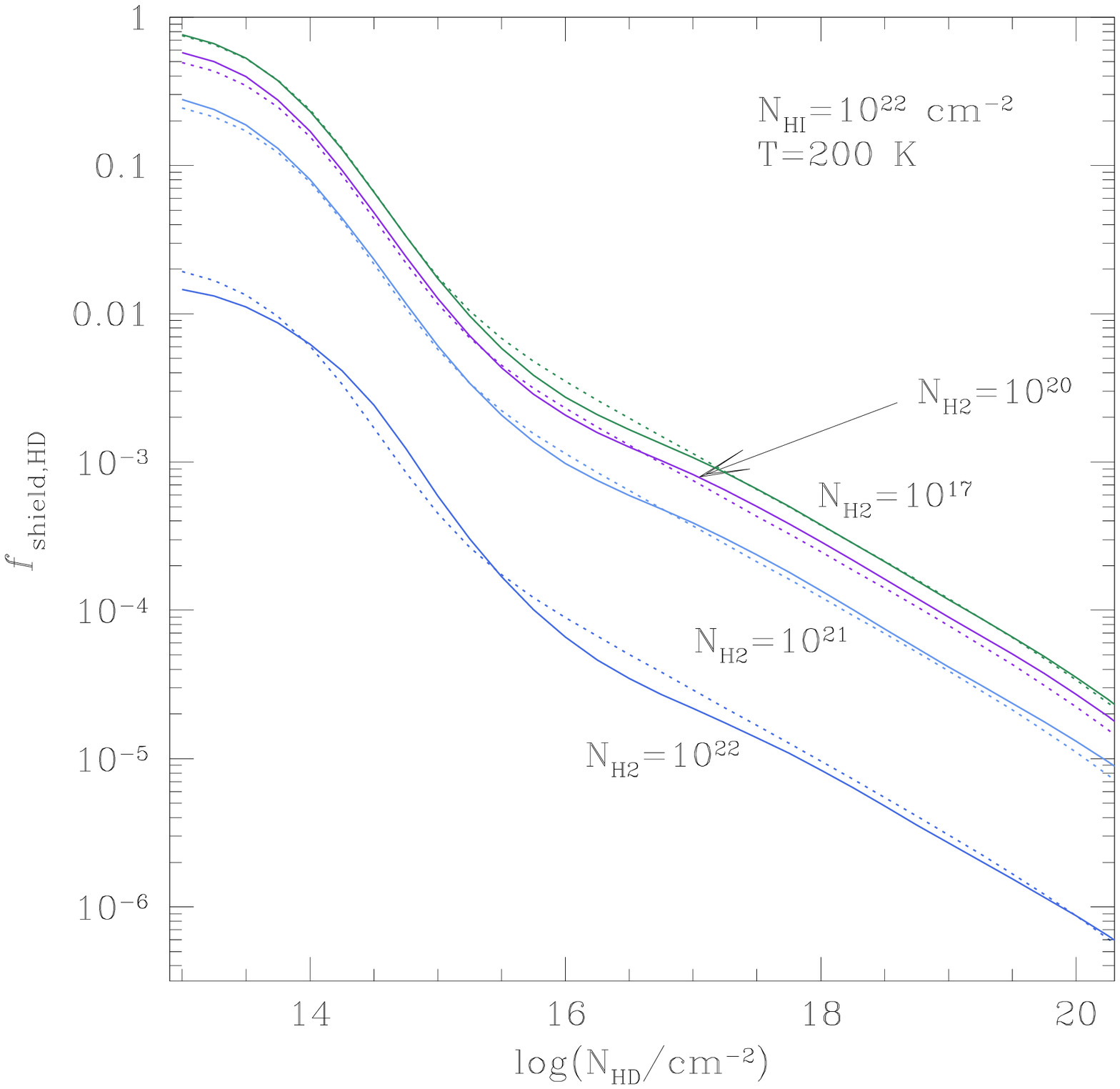}
    \includegraphics[height=2.8in,width=3.4in]{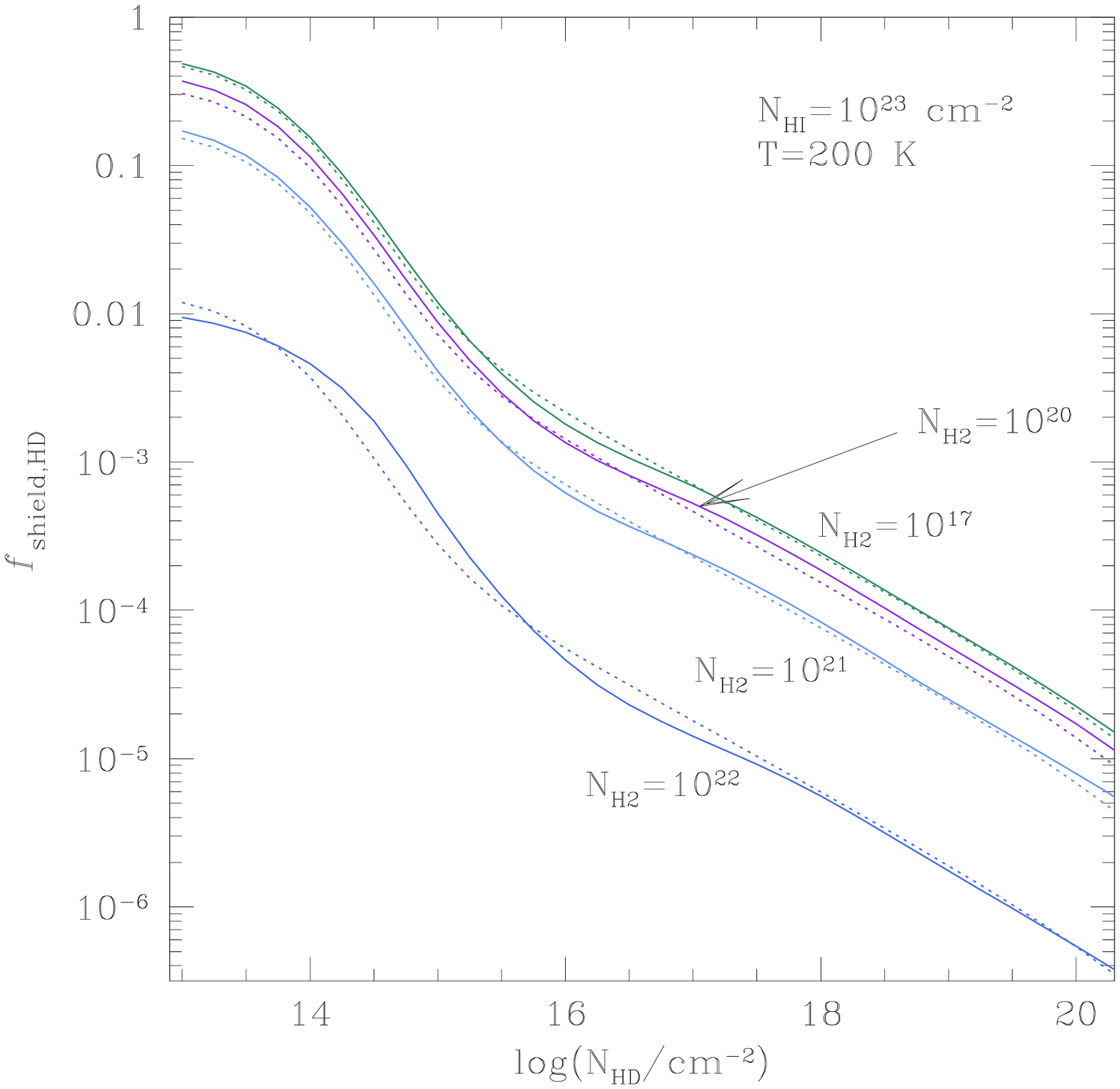}
    \includegraphics[height=2.8in,width=3.4in]{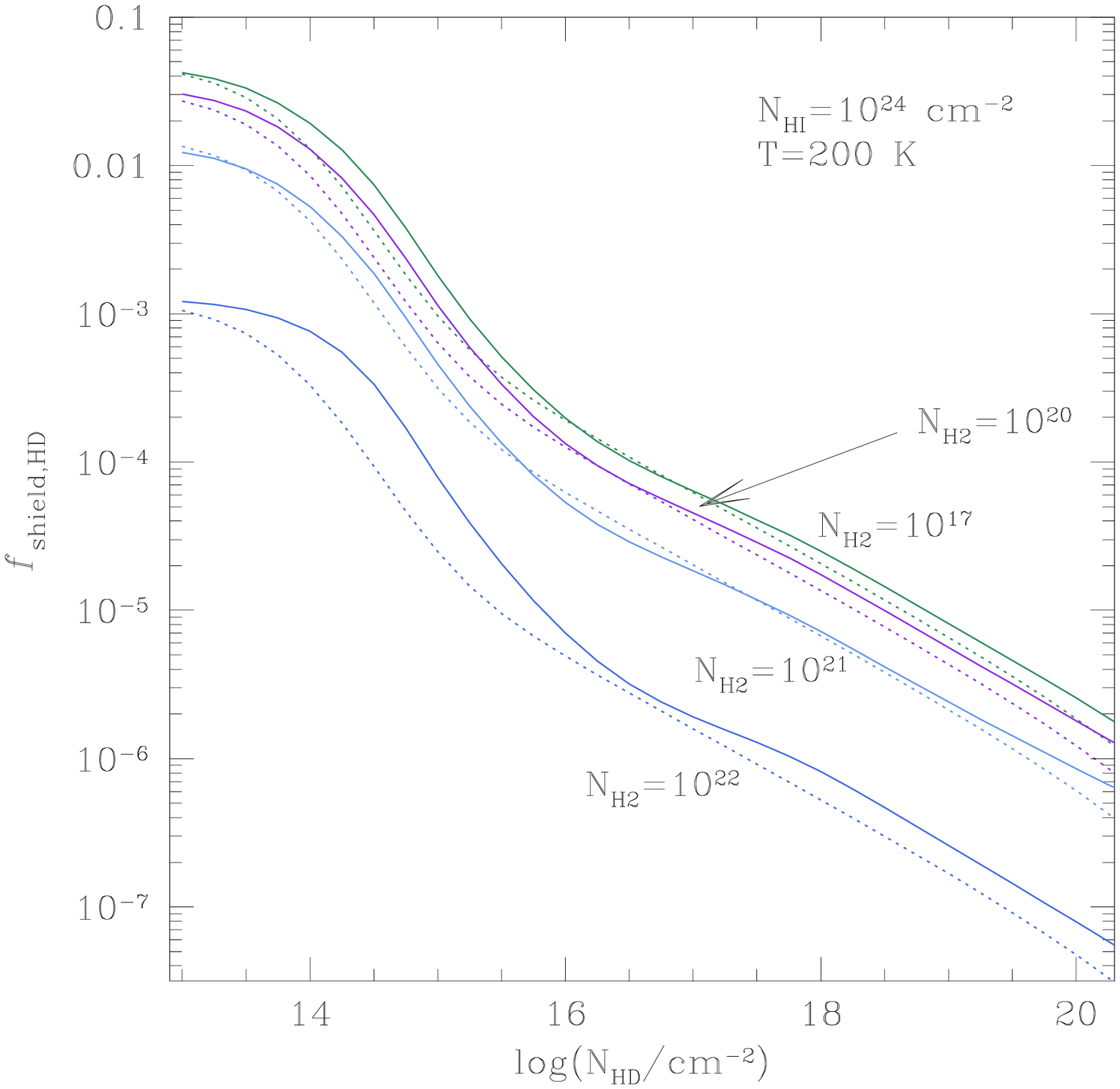}
    \caption{The combined HD shielding factor, including
      self--shielding and shielding by ${\rm H_2}$ and HI.  Several
      combinations of column densities are shown, as labeled, near the
      critical column densities for HI and ${\rm H_2}$ shielding. The
      solid curves show the exact numerical calculations, and the
      dotted curves show the values obtained from a fitting formula
      (equations~\ref{eq:selfshield}, \ref{eq:shieldproduct}, and
      \ref{eq:shield2} and Table~1).}
\label{fig:shieldfits}
\end{figure}

Figure~\ref{fig:shieldfits} shows the results of the exact shielding
factor calculations for a gas temperature of $T=200$K, and compares
these to the analytical fits for a number of combinations of the three
column densities.  The largest deviations are seen in the bottom
panel, for $N_{\rm HI} = 10^{24}~{\rm cm^{-2}}$, and at ${\rm H_2}$
and HD column densities of $N_{\rm H_2} = 10^{22}~{\rm cm^{-2}}$ and
$10^{14}~{\rm cm^{-2}} \lsim N_{\rm HD} \lsim 10^{15.5}~{\rm
cm^{-2}}$.  In this regime, the accuracy of the fitting formulae is
somewhat worse than a factor of two. However, in practice, this column
density combination -- with relatively low $N_{\rm HD}$ and
exceedingly high values of both $N_{\rm H_2}$ and $N_{\rm HI}$ -- does
not occur in our calculations (see Figure~\ref{fig:columns}).

\subsubsection{Shielding of $H_2$ by HI and HD}
\label{subsubsec:H2shield}

The shielding of ${\rm H_2}$ by HD and HI is entirely analogous
to that discussed in the preceding section, so we will limit
the discussion here to a few noteworthy points.
Most importantly, we find that HI shielding of ${\rm H_2}$ 
is nearly identical to HI shielding of HD; accordingly, we
model both with the fitting formula for $f_{\rm shield,HI}
ß(N_{\rm HI})$, given by equation~\ref{eq:shield2} and Table~1.
The explanation for this is similar to that given above; namely,
the relative positions of the ${\rm H_2}$ and HI lines is such that
only the wings of the HI lines shield ${\rm H_2}$ when
the column densities of $N_{\rm HI}$ are sufficiently large
($N_{\rm HI}\gsim 10^{23}$).
Because the ${\rm H_2}$ and HD lines are comparably spaced relative 
to the HI lines, the shielding effect of HI should indeed be
similar for both.

\begin{figure}
    \includegraphics[height=2.8in,width=3.4in]{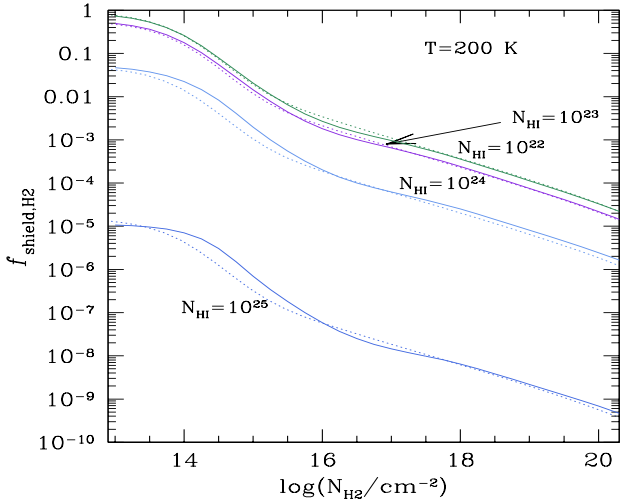}
    \caption{The combined ${\rm H_2}$ shielding factor, including
      self--shielding and shielding by HI.  Several
      combinations of column densities are shown, as labeled, near the
      critical column densities for HI shielding. The
      solid curves show the exact numerical calculations, and the
      dotted curves show the values obtained from a fitting formula
      (equations~\ref{eq:selfshield},\ref{eq:shield2},
      \ref{eq:shieldproductH2} and Table~1).}

\label{fig:shieldfitsH2}
\end{figure}

When the HI column is below the critical level for strong  
shielding of ${\rm H_2}$, we find that a few Lyman lines 
together dominate the dissociation rate, and that the product
$\sigma_\lambda\times f_{\rm diss,\lambda}$ for these lines (at
$\sim$945, $\sim$980, and $\sim$990 \AA) are similar.  
At larger neutral hydrogen columns ($N_{\rm HI}\approx 10^{25}$),
a single Lyman ${\rm H_2}$ line makes the dominant contribution --
by a large margin over all others -- to the dissociation rate. 
Thus, the total shielding factor can again be simply modeled
by a product of the shielding formulae, given in equations~
\ref{eq:selfshield}, \ref{eq:shield2}, and Table~1:
\begin{equation}
f_{\rm shield,H_2} = f_{\rm shield}\left(N_{\rm H_2},T\right)
\times f_1\left(N_{\rm HI}\right)
\label{eq:shieldproductH2}
\end{equation}
The results of
the exact shielding factor and comparison to this
analytical fit are shown in Figure~\ref{fig:shieldfitsH2} 
for a gas temperature of $T=200$K.

In principle, HD can also shield ${\rm H_2}$, but in practice 
this effect will likely always be negligible, as the                             
HD column density is typically dwarfed by those of ${\rm H_2}$ and HI
\footnote {The fractional abundance of HD relative to
that of  ${\rm H_2}$ could exceed the cosmological 
D/H ratio by a large factor, owing to chemical 
fractionation at low temperatures \citep[see, e.g.,][]{GP98}. 
However, it never exceeds $\sim 10^{-2}$ in our models.}. 
Thus, our treatment does not include HD shielding of ${\rm H_2}$. 

\section{Results}
\label{sec:results}

To assess whether HD-cooling can be suppressed by a persistent LW
background, we ran our one-zone models at various different specific
intensities $J_{21}$. Unless stated otherwise, the spectrum of the
radiation is modeled as a black--body with a temperature of $10^5$ K,
approximating the hard spectrum expected to characterize Pop III.1
stars (\citealt{TS00}; \citealt{BKL01}; \citealt{S02}).  For
comparison, in \S~\ref{subsubsec:T4} below, we investigate the effects
of illumination by a cooler blackbody, $T \sim 10^4$K, intended to
represent the softer spectrum of a more typical metal--enriched
stellar population.  These are referred to hereafter as types `T5' and
`T4' respectively (SBH10).

We use a Newton-Raphson scheme to determine the strength of the LW
radiation required to keep the gas temperature greater than a factor
of $\sim 2$ above that reached in the absence of any LW radiation,
$T_{{\rm min},J=0}$, on the timescales described below;
this is referred to hereafter as the critical intensity: $J_{\rm
  crit,HD}$ (in the usual units of $10^{-21}\,{\rm erg \; cm^{-2}\;
  sr^{-1}\; s^{-1} \; Hz^{-1}}$).

\subsection{HD-Cooling in Constant-Density Fossil HII Gas}
\label{subsec:fossils}

The first of the physical scenarios we consider is a ``fossil'' HII
region, which could occur in a patch of the low--density IGM that has
been photo-ionized and heated by a short--lived massive star
\citep[see, e.g.][]{OH03}, or possibly in a denser shell of primordial
gas, compressed by shocks from a supernova (SN).

We are interested in whether gas with such fossil ionization can cool
efficiently in the presence of a LW background and return, via HD
cooling, to a state close to its initial low-entropy state, prior to
the ignition of the ionizing source (or prior to its shock heating).

\begin{figure}
  \includegraphics[height = 3.45in,width=3.3in]{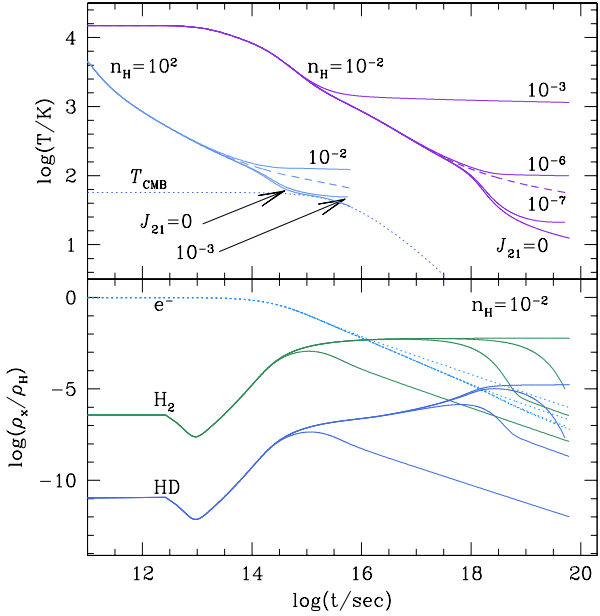}
  \caption{The top panel shows the thermal evolution of constant
    density gas that is initially fully ionized.  The right/left set
    of solid purple/blue curves adopt densities of $10^{-2}$ and
    $10^2~{\rm cm^{-3}}$, appropriate for a flash--ionized patch of
    the IGM and for primordial gas compressed by a SN--shock,
    respectively.  In both cases, we show the evolution when the gas
    is exposed to LW backgrounds of various intensities, as labeled.
    The dashed lines show the temperature reached in deuterium-free
    gas, in the absence of any LW.  The temperature of the CMB, $T{\rm
      _{CMB}} = 2.7 (1+z)$, is shown by the dotted blue curve
    (starting from $z=20$).  The bottom panel shows the fractional
    abundances of electrons (light blue dotted lines), HD, and ${\rm
      H_2}$ (solid blue and green lines respectively) in the
    low--density case.  The fractional abundances are not shown in the
    case of the shock-compressed gas for the sake of clarity, and
    because the patterns they follow are identical to those shown in
    the lower panel.  This figures shows that there is a critical
    flux, $J_{\rm crit,HD}=3.8 \times 10^{-2}(n/10^2~{\rm cm^{-3}})$, 
    that suppresses HD--cooling and prevents the gas from reaching 
    the CMB temperature.}
\label{fig:fossil}
\end{figure}

In a first set of runs, we assume that the number density remains
constant at the low value of $n = 10^{-2}\;{\rm cm^{-3}}$,
characteristic of a slightly over-dense (by $\sim$ a factor of 10)
region of the IGM at $z=10$. We find that in such a rarefied patch,
the gas is not able to cool on a realistic timescale, because the
HD--cooling time is longer than the present age of the universe even
in the absence of any LW background.  
The thermal evolution of the gas in this case is shown by the right
set of (purple) curves in the upper panel of
Figure~\ref{fig:fossil}. The corresponding fractional abundances of
electrons, ${\rm H_2}$, and HD are shown in the lower panel of the
same figure.  The figure extends to a total elapsed time that exceeds
the Hubble time, and shows that there is, technically, a critical flux
($J_{\rm crit,HD}\approx 10^{-6}$) that would suppress the HD--cooling
that would otherwise occur after a few $\times 10^{18}$ seconds. This,
of course, is unphysical, and in practice, the question of HD--cooling 
being suppressed by UV radiation is moot for such low--density
gas. Nevertheless, the figure illustrates the chemical/thermal
behavior, and also provides a useful check on our code (see below).

In the second set of runs, the total particle number density was set
to a higher value of $n = 10^{2}\;{\rm cm^{-3}}$. This is an
unphysically high density for a characteristic ``flash--ionized''
fossil region in the low--density IGM, but may represent primordial
gas compressed by SN shocks.  In the no-flux case, we reproduce the
main result of \citet{JB06}, namely that HD-cooling allows the gas to
reach the temperature of the CMB in a time that is shorter than the
Hubble time.  This gas, however, is optically thin to radiation in the
LW bands, and dissociation of HD (as well as of ${\rm H_2}$) is
efficient. We find that for $J_{21} \gsim 10^{-2}$, the gas cannot
cool to temperatures less than $\sim 200$ K. This is shown in the top
panel of Figure~\ref{fig:fossil} by the left set of (blue) curves. 

The above two cases ($n = 10^{-2}$ and $n = 10^{2}~{\rm cm^{-3}}$) serve to
illustrate an important point (and a check on our code).  All of the
relevant timescales for the system, including the formation timescale
for HD ($t_{\rm form}$), the HD cooling time ($t_{\rm cool}$), and the
(HII+$e\rightarrow$HI) recombination time ($t_{\rm rec}$), scale as
1/{\it n}.  The exception is the photo--dissociation timescale, which
scales with the flux strength, $t_{\rm diss}\propto J_{21}^{-1}$, in
the absence of shielding.  Thus, the history of the system should only
depend on $J_{21}/n$ when the time is rescaled
accordingly.\footnote{The interested reader can find a much more
  detailed discussion of this point for the analogous case of ${\rm
    H_2}$ cooling in \citealt{OH02}.} This simple scaling is evident
by the two sets of (purple and blue) curves in the top panel of
Figure~\ref{fig:fossil}.  Most importantly, there is indeed a critical
flux that prevents the gas from reaching the CMB temperature by HD
cooling; in constant density gas with a high initial electron
fraction, we find the value of this flux is $J_{\rm
  crit,HD}=3.8 \times 10^{-3}(n/10^2~{\rm cm^{-3}})$.

Finally, an interesting question is whether HD--cooling prevented,
for the cases in which $J_{21}$ exceeds the critical value, by direct 
photo--dissociation of HD, or the inability of sufficient HD to
form due to ${\rm H_2}$ photo--dissociation.  To answer this question,
we performed runs in which the ${\rm H_2}$--dissociation was
artificially turned off.  In these runs, we find that the gas is still
able to cool to $T {\rm_{CMB}}$ for $J_{21} \approx 4 \times 10^{-2}$,
illustrating that the LW flux prevents HD--cooling {\em via $H_2$
destruction}, rather than via direct HD--dissociation.  This point has
been discussed by previous authors, e.g. \citet{NU02} showed that a
critical abundance of ${\rm H_2}$, $x_{\rm H_2} \gsim 10^{-3}$ is
required for the gas to reach sufficiently low temperatures for HD to
become the dominant coolant ($T \lsim 150$K), and therefore ${\rm
H_2}$ dissociation can prevent HD--cooling \citep{YOH07}.  
The minimum temperature reached by fossil ionized primordial gas in
the absence of HD, and the dependence of this minimum temperature on
gas density and LW flux was also discussed in detail by \citet{OH03}.

\subsection{HD Cooling in Collapsing Halos}
\label{subsec:collapse}

\subsubsection{HD Cooling in Halos with $T_{\rm vir}>10^4$K}
\label{subsubsec:collapse}

It has been shown that primordial gas in the late stages of runaway
gravitational collapse can reach temperatures close to $T_{\rm CMB}$
via HD-cooling, provided that a large initial ionization fraction
exists (e.g. \citealt{JB06}; see also \citealt{Mach05}).
 
This scenario can be realized in sufficiently massive halos, which are
collisionally ionized upon shock-heating to their virial
temperatures. The post--shock gas can cool faster than it recombines,
leaving a large out-of-equilibrium electron fraction to catalyze both
${\rm H_2}$ and HD formation \citep[e.g.][]{SK87,Susa+98,OH02}.  It
may also be the case that ``pre-ionized'' halos exist within fossil
HII regions, which will undergo a phase of efficient HD--cooling upon
collapse \citep{JB06}. This scenario, however, is less plausible:
a halo large enough to remain bound once photo-heated (to $T\gsim
10^4$K) may be difficult to completely ionize, as the large HI column
densities will lead to non-negligible HI self-shielding 
\citep{Dijkstra+04}; flash--ionization by a single short--lived star
(required to allow subsequent recombination and cooling) is even less
likely.
   
\begin{figure}
    \includegraphics[height=3.45in,width=3.3in]{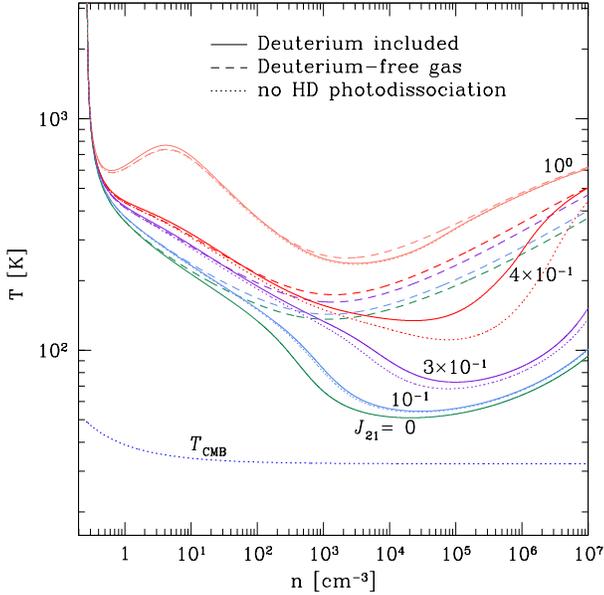}
    \caption{Thermal evolution of initially ionized gas, collapsing in
      a massive halo ($T_{\rm vir} \gsim 10^4$K) exposed to
      LW backgrounds in the range near the critical value,
      $J_{\rm crit,HD} = 3.6 \times 10^{-1}$ (solid curves). The incident
      spectrum is that of a blackbody with a temperature of $10^5$K,
      characteristic of massive Pop III.1 stars.  The turnaround
      redshift is set to $z = 17$ and the temperature is initialized
      $T \approx 10^4$K. The dashed curves show, for comparison, the
      temperature evolution in deuterium-free gas, exposed to the same
      LW fluxes.  The dotted curves show the thermal evolution
      when HD-dissociation is artificially switched off, for the
      same values of $J_{21}$, as labeled. This illustrates 
      that, except in a small range of flux intensities,
      the destruction of ${\rm H_2}$ by LW photons,
      rather than direct HD photo--dissociation, is the primary 
      factor determining the minimum temperature reached in the gas.
      The temperature of the CMB is shown by the blue dotted curve.}
\label{fig:collapse}
\end{figure}

Regardless of the nature of the initial ionization,
Figure~\ref{fig:collapse} shows the thermal evolution of the
collapsing gas exposed to LW backgrounds of various intensities.  The
initial number density is set to the characteristic baryon density in
halos upon virialization,
\begin{equation} 
n \simeq 0.3\; {\rm cm^{-3} \left(\frac{1+z_{vir}}{21}\right)^3},
\end{equation}
and the gas begins cooling from the temperature $T\approx 10^4$K
(quickly established either by a period of photo--heating, or by
shock--heating to near the virial temperature, accompanied by rapid HI
cooling).

Of primary importance in this case -- as opposed to the fossil gas
discussed in the previous section -- are the large column densities of
HD and ${\rm H_2}$ that build up and shield both populations against
the LW background.  Furthermore, the collapse itself leads to more
efficient formation of both molecules because the formation timescale,
as mentioned above, scales as $t_{\rm form} \propto
1/n$. Consequently, the critical intensity should be larger than that
found for the low-density fossil gas.  This is indeed borne out by our
results; nevertheless, as the comparison between the solid and the
dashed curves (the latter representing deuterium--free gas) in
Figure~\ref{fig:collapse} shows, the effect of HD--cooling is still
almost entirely erased for the relatively low values of $J\gsim 1$.
The critical intensity in this case, as defined above, is found to 
be $J_{\rm crit} = 3.6\times 10^{-1}$.

{\em This threshold value is most notable for being approximately
five orders of magnitude lower than the critical flux required 
to completely suppress ${H_2}$--cooling in the same halos.}  
As shown in SBH10, the latter critical flux in halos with 
$T_{\rm vir}\gsim 10^4$K is $J_{\rm crit,H_2} \gsim 10^4$.  This large
critical flux corresponds to the value that results in an ${\rm
H_2}$--photo--dissociation rate that matches the ${\rm H_2}$ formation
rate, at the critical density of $n\sim 10^4~{\rm cm^{-3}}$ of ${\rm
H_2}$ (see the earlier work by O01 for a detailed discussion
of the physics determining $J_{\rm crit,H_2}$ in primordial gas
without ionization/shock--heating).  This $J_{\rm crit,H_2} \approx
10^4$ separates gas in which ${\rm H_2}$--cooling is {\em fully}
suppressed (with the gas temperature remaining near $\sim 10^4$K) and
halos in which ${\rm H_2}$ cooling significantly lowers the
temperature.  A point that was also found (but not emphasized) by
SBH10 (and also by O01) is that even for $J_{21}$ well
below $J_{\rm crit,H_2}$, the minimum temperature to which the gas can
cool via ${\rm H_2}$ can be significantly elevated. This is also
clearly visible in the deuterium--free runs in
Figure~\ref{fig:collapse}: the minimum temperature is $\sim 150$K for
$J_{21}=0$, but is elevated to $\sim 300$K already for $J_{21}=1$.  

The behaviour of the gas, and the reason for the elevated temperature,
can be described as follows (see a detailed discussion in the
constant--density case in \citealt{OH02}).  Starting from $T\approx
10^4$K, the gas initially cools via ${\rm H_2}$ and recombines on
time--scales much shorter than either the photo--dissociation or the
free--fall timescale.  However, when the temperature is lowered to a
$J$--dependent critical value of a few$\times10^3$K, ${\rm
H_2}$--dissociation becomes important, limiting the ${\rm H_2}$
abundance, and reducing the cooling. The cooling time eventually
becomes comparable to the free--fall time, resulting in the sharp turn
away from the nearly vertical directions of the $n-T$ curves at the
initial density in Figure~\ref{fig:collapse}. For higher fluxes, this
subsequently results in an elevated gas temperature (at fixed density).
Eventually, the compressional heating rate becomes equal to the ${\rm
H_2}$--cooling rate, setting the temperature minimum.  It is worth
noting that for sub--critical values of $J_{21} \lsim J_{\rm crit,HD}$, HD
is able to cool the gas to temperatures below $150$K, but the LW
background still has the subtle effect of raising the minimum
temperature to which the gas can cool via ${\rm HD}$.

As in the constant density case, an interesting question is whether
ultimately the HD--cooling is controlled by direct photo--dissociation
of HD, or by ${\rm H_2}$--dissociation.  To answer this
question, we repeated the runs shown in Figure~\ref{fig:collapse}
under the same conditions, but with HD dissociation artificially
switched off (the results are shown by the dotted lines in
Figure~\ref{fig:collapse}).  In this case,
we find again that (except in a narrow range of LW intensities
near $J_{21} \sim 4\times 10^{-1}$), direct photo--dissociation 
of HD does not determine the minimum temperature reached in the gas.
Rather, it is the diminished abundance of ${\rm H_2}$ in the presence
of the LW background that regulates the abundance and thereby the
cooling efficiency of HD.

\begin{figure}
    \includegraphics[height=3.45in,width=3.3in]{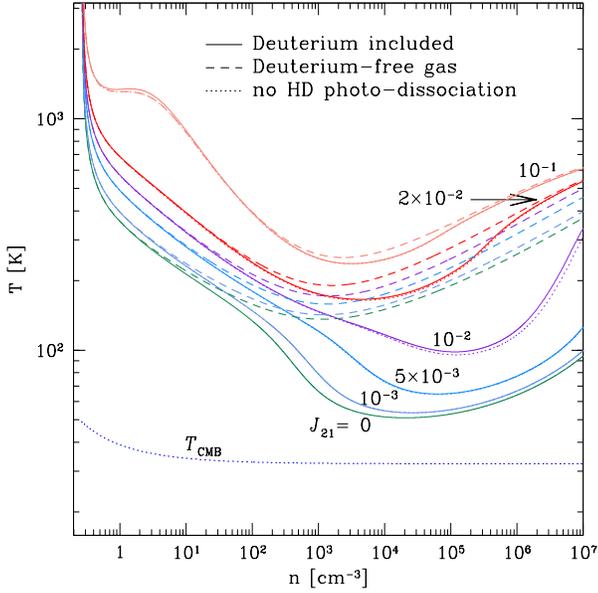}
    \caption{Thermal evolution of gas collapsing in a massive halo,
      exposed to different LW backgrounds, as in
      Figure~\ref{fig:collapse}, except the incident spectrum is that
      of a softer blackbody with temperature $T=10^4$K, representing a
      more typical metal-enriched stellar population.  This softer
      spectrum contains many more photons down to the
      photo--dissociation threshold of ${\rm H^-}$ at 0.76eV. The
      enhanced rate of ${\rm H^-}$ photo--dissociation reduces the
      critical flux that prevents HD--cooling by more than an order of
      magnitude compared to the harder spectrum, to $J_{\rm crit,HD}
      = 10^{-2}$.}
\label{fig:collapseT4}
\end{figure}

We have also investigated the thermal evolution of the gas when 
${\rm H_2}$ is artificially prevented from dissociating (not shown), 
but the physical set-up is otherwise analogous to the runs
(in which deuterium is included) shown in Figure~\ref{fig:collapse}. 
This ``academic'' exercise is useful in order to determine the 
critical flux that would prevent HD-cooling entirely by direct 
HD photo--dissociation.  In the analogous case for
${\rm H_2}$, as mentioned above, $J_{\rm crit, H_2}$ is traditionally
defined as the specific intensity capable of {\em completely}
suppressing ${\rm H_2}$-cooling, thereby preventing the gas from
falling below the temperatures reached by atomic line cooling, $T \sim
8 \times 10^{3}$K (O01, OSH08, SBH10).  
We find that for $J \gsim 6\times 10^4$, the cooling history of the 
gas is nearly identical to that of deuterium-free gas in the 
absence of a LW background. Thus, the intensity required to 
fully suppress HD--cooling by direct HD--dissociation is  
comparable to that of ${\rm H_2}$ (the latter was found by the
latest studies (SBH10) to be $\sim 1.2 \times 10^4$, but we find a
factor of $\sim$ five greater critical value for ${\rm H_2}$; 
see below). In fact, this is not surprising, given that the 
photo--dissociation timescales ($t_{\rm diss} = [k_{\rm diss} 
(N=0) \times f_{\rm shield}]^{-1}$) are very similar for the 
two molecules.

\subsubsection{$T_{\rm vir}>10^4$K Halos Illuminated by `T4' Radiation}
\label{subsubsec:T4}

Up to this point, we have considered only incident radiation with the
hard spectrum expected to characterize Pop III.1 stars (`T5'). In the
case of a collapsing halo, it is reasonable to ask about the effects
of irradiation by the more the typical stellar spectrum (`T4'), on
HD-cooling.

Figure~\ref{fig:collapseT4} shows the temperature evolution in the gas
irradiated by a T4 spectrum of various intensities. It is clear that
HD is considerably more fragile in the presence of this softer
spectrum, withstanding a LW flux no greater than a feeble $J_{\rm
crit,HD} = 10^{-2}$.  This is not surprising in the light
of studies that have found the same effect for ${\rm H_2}$ (O01;
OSH08; SBH10), namely, that ${\rm H_2}$--cooling is much more
effectively suppressed by the T4 type spectrum. The reason is the
diminished abundance of hydride (${\rm H^-}$), an intermediary in 
the formation of both ${\rm H_2}$ and HD. Hydride, 
whose ionization threshold is 0.76eV, is more efficiently 
photo-dissociated by the softer spectral type (O01, SBH10).  
This again is a manifestation of how an external radiation field can 
regulate HD-cooling {\em via} the destruction of an intermediate 
in its formation pathway.

\subsubsection{HD Cooling in $T_{\rm vir}<10^4$K Halos}
\label{subsubsec:minihalos}

\begin{figure}
    \includegraphics[height=3.45in,width=3.3in]{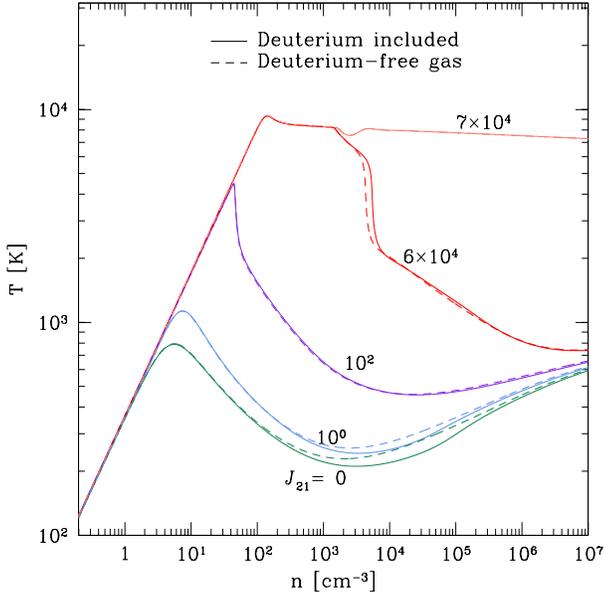}
    \caption{Thermal evolution of gas collapsing in a halo exposed to
      LW radiation (with a T5 spectrum) in Figure~\ref{fig:collapse},
      except that the initial gas temperature is assumed to be much
      lower, 21K.  This is relevant to minihalos ($T_{\rm vir} \lsim
      10^4$K), or to the case when there are no strong shocks able to
      collisionally excite or ionize the gas.}
\label{fig:minihalo}
\end{figure}

It has long been known that pristine gas in the first minihalos cannot
form sufficient ${\rm H_2}$ to cool below a few hundred Kelvin
\citep[e.g.][]{HTL96}.  Because free electrons act to catalyze the
formation of HD, as well as ${\rm H_2}$, it is not surprising that HD
abundances remain too low in such halos to play a significant role in
cooling \citep[e.g.][and references therein]{JB06}. This is borne 
out by our results, shown in Figure~\ref{fig:minihalo}, for a 
halo that begins collapsing from a temperature of $T \approx 20$K at 
the turnaround redshift, $z = 17$, and is illuminated by a `T5' spectrum.  
This is the same configuration as in Figure~\ref{fig:collapse}, 
except that the initial temperature is assumed to be low (i.e., lacking 
any strong shocks able to collisionally excite or ionize the gas).

We include a discussion of this scenario for the purpose of
highlighting a few noteworthy points. First, we find a critical flux
for full ${\rm H_2}$--dissociation in this case (as it is
traditionally defined, see \S~\ref{subsubsec:collapse} above) of
$J_{\rm crit,H_2} = 6.1 \times 10^{4}$, which is a factor of $\sim$ 
five greater than that found in the recent study by SBH10.  
This difference owes to our use of the analytic fit for 
${\rm H_2}$ self--shielding (equation~\ref{eq:selfshield}), 
which assumes all molecules are initially the ground state, while 
SBH10 used the shield factor fit (equation 36) from DB96, which
gives a very good approximation to the self--shielding when the
${\rm H_2}$ roto--vibrational levels reach LTE populations.
This illustrates an important point that will be discussed in 
greater depth in ~\S\ref{sec:discussion}; namely, self--shielding 
is {\em less} effective when higher ro-vibrational states of the 
molecule are populated (see more discussion in \S~\ref{sec:discussion} below).
It is also worth noting that we find $J_{\rm crit,H_2} = 
6 \times 10^{3}$ -- a factor of 2 {\em lower} than that found by SBH10 --
if the self-shielding is modeled instead with the more accurate 
formula provided by DB96 (equation 37), rather than the power-law 
fit (equation 36) used by SBH10. In general, the ``real'' LW intensity 
required to kill ${\rm H_2}$-cooling entirely will depend 
on the detailed ro-vibrational distribution of ${\rm H_2}$ 
molecules. The values of $J_{\rm crit,H_2}$ found here 
using the ground-state {\it vs} LTE shielding treatments serve to bracket 
the range for the true critical threshold, with the former approximating
the upper limit on $J_{\rm crit,H_2}$. The 
threshold flux in this case is of particular interest as it has important
implications for the number of halos that remain ${\rm H_2}$--poor,
and thus for the abundance of halos collapsing directly to massive
seed black holes, because such halos probe the rare, exponential tail
of the fluctuating UV background \citep{Dijkstra+08}. (The interested
reader is encouraged to see SBH10 for a detailed discussion of these
issues.)

Second, while there is a clear bifurcation in the cooling history of
the gas around the $J_{\rm crit,H_2}$ discussed above, we note that it
does not cool to the low temperatures required for HD-cooling to
become important ($T \sim 150$K) even when the intensity is well below
this threshold (indeed, even in the absence of any LW background).
Thus, again, ${\rm H_2}$ abundance plays the primary role in regulating
HD-cooling, though in this case the result is simpler: HD--cooling never
becomes important because ${\rm H_2}$--cooling is never strong enough 
for the gas temperature to fall below $\approx 200$K.

Finally, in this case HI shielding serves to increase the threshold
LW flux by $\sim 40\%$ above that which ${\rm H_2}$ could otherwise withstand.
By contrast, $J_{\rm crit,HD}$ found in the two preceding sections
is not sensitive to the effect of ${\rm H_2}$ shielding by HI, in spite 
of its strong dependence on the ${\rm H_2}$ abundance. This is because, 
in the models of $T_{\rm vir} \gsim 10^{4}$K halos, the HI column density 
does not reach the critical value above which it effectively shields 
${\rm H_2}$ ($N_{\rm HI} \gsim 10^{23}~{\rm cm^{-2}}$) until very late in the 
collapse (i.e., once the particle density has reached 
$n \gsim 10^{5.5}~{\rm cm^{-3}}$). However, sufficiently high column 
densities of HI do build up earlier in the high-$J_{21}$ runs of
$T_{\rm vir} < 10^4$K halos, resulting in the modest increase 
in $J_{\rm crit,H_2}$ as quoted above.

The results for the halo illuminated by a T4 spectrum are omitted here
because they do not significantly differ from those described above
for the T5 case, except that ${\rm H_2}$-cooling is disabled by a much
lower LW flux, as shown already by O01.

\section{Discussion}
\label{sec:discussion}

The most significant result found in this paper is that a UV flux can
prevent HD--cooling from lowering the gas temperature to near that of the CMB.
In particular the threshold value in collapsing halos, $J_{\rm crit,HD} \sim 4
\times 10^{-1}$, is approximately five orders of magnitude lower than the critical flux
required to completely suppress ${\rm H_2}$--cooling in the same halos.
As explained in \S~\ref{subsubsec:collapse}, this large difference
arises because even for $J_{21}$ well below $J_{\rm crit,H_2}$, the minimum
temperature to which the gas can cool via ${\rm H_2}$ is significantly
elevated, so that HD formation and cooling is not activated.

\noindent{\em What are the cosmological implications if a LW
background prevents halos from cooling via HD line emission?}  The
critical flux we find should be compared to the level of the LW
background $J_{\rm bg}$ expected at the redshifts we consider.  For
reference, an estimate of the background as a function of redshift is
provided by requiring the number of UV photons produced by stars to be
sufficient to reionize the IGM.  Such an estimate gives
\citep[e.g.][]{BL03}
\begin{equation}
J_{\rm bg} \approx 4 \left(\frac{N_{\gamma}}{10} \right) \left(
\frac{f_{\rm esc}}{100\%} \right)^{-1} \left( \frac{1+{\rm z}}{11}
\right)^3.
\end{equation}
Here the number of UV photons required to ionize hydrogen is
commonly taken to be $N_{\gamma} \sim 10$, and the escape fraction
$f_{\rm esc}$ is the fraction of ionizing photons ($13.6{\rm eV}$) escaping
from galaxies at high redshifts, which is expected to be close to
unity (\citealt{WAN04}; see \citealt{FS10} for a recent discussion of
the relevant $f_{\rm esc}$ and for references to earlier works).  From
this equation, we find that the mean UV background at the time of
reionization exceeds the critical flux $J_{\rm crit,HD}$ by nearly
an order of magnitude.  While the radiation background will inevitably have
spatial fluctuations, this implies that most halos collapsing in the
early IGM, prior to reionization at $z \sim 10-20$, would be exposed
to a super-critical flux and thus not able to cool below $T \sim 200$K.
As a result, the emergence of low--mass PopIII.2 stars (or stars
comparable in mass to those formed in the low-redshift universe) would
be postponed until supernovae polluted the IGM with heavy elements,
and metal-line cooling subsequently enabled gas clouds to reach
temperatures near $T_{\rm CMB}$.
 
There are several issues that could be important for the above
conclusion, which we have glossed over in the discussion of our
models for molecular shielding and one-zone spherical collapse. 
We next discuss some of these.

\noindent{\em Uncertainty in the Column Density of the Collapsing
Region.} In our calculations, we have taken the diameter of the
collapsing region to be of the order of the Jeans length. However, the
effective size and column density will depend sensitively on the
dynamical properties of the system, including bulk motions and
internal velocity gradients in the gas, departures from spherical
symmetry, etc.

In order to address this uncertainty quantitatively, we have performed
two additional sets of runs for a halo with $T_{\rm vir} \gsim 10^4$K
(analogous to those in \S~\ref{subsubsec:collapse}). In the first, we
increase the assumed size of the collapsing region by a factor of ten 
(i.e. to ten times the Jeans length $\lambda_{J}$). This increases the 
column densities by the same factor, and accordingly, $J_{\rm crit}$ is larger by a factor 
of $\sim 3$ than the original result, due to more 
efficient self-shielding of ${\rm H_2}$ and -- to a lesser extent -- HD.
Next, we investigated the cooling properties of a smaller collapsing 
core, with the assumed size reduced by a factor of ten, to  $0.1\lambda_{J}$. In this case, a new effect arises. 
Namely, for the values of $J_{21}$ at which the gas can (just) cool to around $T \sim
200$K, we find that direct HD dissociation is the dominant factor
regulating the minimum temperature ultimately reached.  In particular,
switching HD dissociation off by hand in these cases decreases the gas
temperature by a factor of $\sim 1.5$ at high number densities ($n
\gsim 10^4 \; {\rm cm^{-3}}$) for $J = 10^{-1}$.  When the flux is
weaker than this, artificially disabling the HD-dissociation has
little effect; for these low fluxes, however, the suppression of
HD--cooling is modest to begin with. As expected, the critical flux is
decreased in this case because the smaller column densities leave both 
${\rm H_2}$ and HD more susceptible to dissociation. In particular,
we find a critical value of $J_{\rm crit}\sim 10^{-1}$.

\noindent {\em The Impact of 3-D Gas Dynamics on Self-Shielding.} It
is important to note that a full treatment of the three- dimensional
dynamics of the system and the complexities inherent in radiative
transfer is needed to solve the shielding problem exactly. Our
calculation is based on a model of a uniform slab of gas with no
internal velocity (or temperature) gradients.
This is likely to be a poor approximation for a region undergoing
runaway gravitational collapse, in which high gas velocities can
produce significant Doppler shifts of the LW absorption bands of ${\rm
H_2}$ and HD. In general, we expect this to {\em reduce} the effective
column densities, and the importance of shielding, compared to our
calculations.  This correction can be mitigated by the broadening of
absorption lines at high column densities, which leads to line widths
that are much larger than the average Doppler shift. In general,
however, taking into account the possibility of Doppler shifts leads
to the conclusion that our self-shielding results, and in turn the
values quoted for $J_{\rm crit,HD}$, are {\em upper limits} for
both. This strengthens our argument above, namely, that most
collapsing halos will see a super-critical flux.

\noindent{\em Resonant Scattering of Incident LW Photons.}
An additional effect ignored by our self-shielding calculation is that
after absorbing a LW photon, a fraction of molecules will decay
directly back to the original ground state, as opposed to cascading
through a series of lower energy decays as we implicitly assume.
These photons are thus not eliminated, making the background flux
stronger than we calculate. However, \citet{GB01} estimate that in the
case of ${\rm H_2}$, such resonant scattering constitutes only a small
fraction, 4-8\%, of all LW absorption events (depending on the initial
level populations; in particular, on the ortho/para ratio).  Hence
this is a minor effect, which again makes our conclusions about the
suppression of HD--cooling conservative.

\noindent{\em Molecular Level Populations: Implications for
  Photo--dissociation Rates and the Critical Flux.} As described in
\S\ref{subsec:opthin}, our fiducial self-shielding calculations 
for HD and ${\rm H_2}$ assume that all molecules are initially in the
ro-vibrational and electronic ground states. In reality, molecules
will occupy higher ro-vibrational states due to collisional excitation, 
and this can significantly {\em increase} the rates of
photo--dissociation. In particular, we find that self--shielding is
less effective (i.e. the shield factor is larger by a factor of a
$\sim$ few) if we repeat our calculations, assuming  LTE population distributions 
in the rotational levels ($J \neq 0$) within the ground electronic and vibrational
state.  (Data for the HD and ${\rm H_2}$ energy levels were taken from
\citealt{ARV82} and \citealt{D84} respectively.)  Other studies have
also noted that populations in higher vibrational levels ($v \neq
0$) can significantly increase the rates of photo--dissociation
\citep[e.g.][and references therein]{GJ07}.  Thus, if the effects of
collisional excitation are taken into account, $J_{\rm crit,HD/H_2}$
could be significantly lower than found above. The ro-vibrational
distribution will be somewhat better approximated by the ground state
model up to the critical densities ($n \gsim 10^4 {\rm cm^{-3}}$ for
${\rm H_2}$ and $n \gsim 10^6 {\rm cm^{-3}}$ for HD).  In our case,
the fate of the collapsing clouds -- whether it will ultimately cool
to temperatures low enough for HD--cooling to become significant -- is
determined in the regime of somewhat lower gas densities, below those
at which equilibrium ${\rm H_2}$ populations are established, so that
$J_{\rm crit}$ values are likely closer to the upper end of the range.

\noindent {\em Uncertainties in the Gas Phase Chemistry.}
The preceding discussion has focused on
various uncertainties associated with photo--dissociation
rates of ${\rm H_2}$ and HD; however, 
the accuracy of any estimates of $J_{\rm crit,HD}$ can 
only be as good as the accuracy of the underlying chemical 
rate coefficients. Considerable attention 
has been dedicated to uncertainties in both ${\rm H_2}$ 
and HD chemistry and cooling, (e.g. \citealt{SKHS04}, 
\citealt{GSJ06}, \citealt{G07}, \citealt{GA08}); accordingly,
we restrict the discussion here to focus on two examples of
thermal rate coefficients that have recently been updated, and how 
their revised values affect our results.  

Two reaction rates that are crucial for determining 
the abundance of ${\rm H_2}$, particularly in gas with a
large initial ionization fraction, have been estimated to be 
uncertain by up to an order of magnitude \citep[e.g.][]{GA08}. 
These are the associative detachment channel for 
${\rm H_2}$ formation, and the mutual neutralization
of hydride and protons (reactions 10 and 13 respectively in 
the compilation by SBH10): 
\begin{equation}
{\rm H + H^- \rightarrow H_2  + e^-}
\label{eq:mutualneutralization}
\end{equation}
\begin{equation}
{\rm H^+ + H^- \rightarrow H + H}.
\label{eq:AD}
\end{equation}
Both have been revisited recently and new thermal rate coefficients
have been provided for reaction (10) by \citet{Savin+10}
and for reaction (13) by \citet{Stenrup+09}.
As noted above, the formation of HD occurs primarily
via the reaction pathway shown in equations~\ref{eq:reaction1}--
\ref{eq:reaction3}, so in general, the fractional abundance of HD
is proportional to that of ${\rm H_2}$, justifying the emphasis here
on uncertainties associated with the formation of ${\rm H_2}$.
In order to quantify how these recently updated rate
coefficients impact our results, we have performed additional 
runs for each of the physical scenarios discussed in 
\S~\ref{sec:results}. The uncertainties these introduce into 
estimates of $J_{\rm crit,HD}$ and the minimum temperature, 
$T_{\rm min}$, reached when $J_{21} = 0$, are summarized below.

We examine three published rate coefficients for the mutual
neutralization reaction; the largest of these at all temperatures is 
that provided by {GP98}, hereafter $k_{\rm 13b}$.  The rate used 
in our fiducial model ($k_{13}$), originally provided by \citet{DL87}, 
is smaller than the others by up to $\sim$ an order of magnitude 
for $T\gsim 10^{3}$K, and by a factor of several
at lower temperatures. \footnote{\citet{GSJ06} have suggested 
that this rate is in fact erroneously small, perhaps due to 
typographical errors in the source. Nonetheless, it is
widely used in studies of star formation in metal--free gas.} 
The value of the newest thermal rate coefficient, 
given by \citet{Stenrup+09}, $k_{\rm 13c}$, lies between those of 
$k_{13}$ and $k_{\rm 13b}$ at all temperatures in the regime we study. 
Using $k_{\rm 13b}$, we find the critical flux, $J_{\rm crit,HD}$
is a factor of $\sim 2$ lower than was found in each of the 
physical scenarios we have studied (see \S~\ref{sec:results}; 
note that this does not apply to the model of halos with 
$T_{\rm vir} < 10^4$K), and $T_{\rm min}$ is $\sim 30 \%$ 
greater in the spherical collapse models. This is easily 
explained: when a large ionization fraction exists, 
reaction (13) competes with reaction (10) for the common
reactant, ${\rm H^-}$; thus, adopting a larger rate coefficient 
for reaction (13) leads to diminished abundance of, and less 
robust cooling by ${\rm H_2}$. Using the newest rate,
$k_{\rm 13c}$, we find the critical flux is $\sim 30-40\%$ 
smaller than its original value for each physical scenario, 
and the minimum temperature is elevated $\sim 20\%$ above 
that found previously in the spherical collapse models. 
(Note that the minimum temperature reached in the fossil HII
region models does not depend on which rate coefficient is used 
for reaction 13.)

The effect of implementing the new rate for 
associative detachment is less dramatic; note, however, that the
uncertainty associated with this rate coefficient has been 
reduced thanks to the recent study by \citet{Savin+10}.
Using the systematic uncertainty given by the authors, 
we implement the thermal rate at the $\pm~ 1\sigma$ levels,
and find the following ranges for the critical flux.
In the fossil HII region: $J_{\rm crit,HD} = 
3.85 \pm 0.15 ~ \times~ 10^{-3} [n/10^2]$. 
In $T_{\rm vir} > 10^4$K halos: 
$J_{\rm crit,HD} = 3.7 \pm 0.2 ~\times~ 10^{-1}$ and 
$1.2 \pm 0.3 ~\times~ 10^{-2}$
when illuminated by T5 and T4 spectral types respectively. 
For a thorough discussion of how this new 
rate coefficient compares to previous calculations and its impact 
on ${\rm H_2}$ chemistry, the reader is referred to \citet{Savin+10}.

Two final notes on the chemical network are on order: 
first, we have not included in this study the formation of ${\rm H_2}$ 
via a three-body reaction, by which metal--free gas can
become fully molecular at $n \gsim 10^8~{\rm cm^{-3}}$,
because our models do not include these high-density regimes.
Lastly, as mentioned above, the primordial value of D/H varies in the
literature by a factor of $\sim 2$, and this in itself may have
consequences for the role of HD--cooling in metal--free gas. 

\noindent {\em Fragmentation and Characteristic Protostellar Masses.}
Finally, we emphasize that all our conclusions in this paper are based
on the thermal history of a gas cloud, and how this is affected by a
UV flux.  In order to make realistic predictions for the
fragmentation, and the ultimate sizes of stars forming in a collapsing
halo, fully 3-D hydrodynamical simulations would be required. While
robust HD--line cooling (or lack thereof) could have a notable impact
on the characteristic stellar masses in the earliest dwarf galaxies,
the process of fragmentation is yet to be fully understood, and thus
physical ingredients such as the minimum gas temperature may not
directly translate into the actual mass of the protostar that
ultimately forms (see, e.g., \citealt{Clark+10} for recent results on
fragmentation in metal--free gas, and in particular the importance of
turbulence).

\section{Summary}
\label{sec:conclusions}

We have demonstrated that HD--cooling in primordial gas can be
suppressed by a relatively weak external LW background, with
an intensity on the order of $J_{21} \sim 10^{-3}(n/10^{-2}~{\rm
cm^{-3}})$ in constant-density ``fossil ionized'' gas, and $J_{21}\sim 
10^{-1}$ in shock-- or photo--ionized gas collapsing into 
halos with virial temperatures greater than $\sim 10^4$K.  These 
critical intensities are lower than the expected mean UV background at
$z\sim 10-20$, suggesting that HD-cooling is likely unimportant in
most proto-galaxies forming near and just prior to the epoch of
reionization.  We conclude that an ``HD-mode''
of star formation was not as prevalent as previously thought.  

On a more technical note: we have also found that the negative
feedback of the LW background is mediated via the abundance of
molecular hydrogen, which is dissociated by the same radiation in its
Lyman and Werner bands. Direct HD photo--dissociation is comparatively
less important, although we find that in regimes of less
effective self--shielding, it can regulate the minimum
temperature of the gas.  Finally, we have provided fitting formulae
for the effects of HD and ${\rm H_2}$ self-shielding, shielding of
both species by HI, and shielding of HD by ${\rm H_2}$, 
which we hope will be useful in other future studies.

\section{Acknowledgments}

We would like to thank Volker Bromm, Greg Bryan, Simon Glover and
Daniel Savin for useful discussions.  This work was supported by the
Pol\'anyi Program of the Hungarian National Office for Research and
Technology (NKTH).

\bibliography{HD}

\end{document}